\documentclass[twocolumn,showpacs,preprintnumbers,amsmath,amssymb,prb]{revtex4}

\usepackage{graphicx,graphics}
\usepackage{bm}

\begin{document}
\title{The radial breathing-like mode of the collapsed Single-walled
carbon nanotube bundle under hydrostatic pressure}

\author{Gang Wu}
\altaffiliation[Present address: ]{Department of Physics, National
University of Singapore, Singapore 117542}
\email{wugaxp@gmail.com}

\author{Xiaoping Yang}
\altaffiliation[Also at: ]{Department of Physics, Huainan Normal
University, Huainan, Anhui 232001, P. R. China}

\author{Jinming Dong}
\email[Corresponding author. Email address: ]{jdong@nju.edu.cn}

\affiliation{National Laboratory of Solid State Microstructures
and Department of Physics, Nanjing University, Nanjing 210093, P.
R. China}

\begin{abstract}
Using the first principles calculations we have studied the
vibrational modes and Raman spectra of a (10, 10) single-walled
carbon nanotube (SWNT) bundle under hydrostatic pressure. Detailed
analysis shows that the original radial breathing mode (RBM) of
the SWNT bundle disappears after the structural phase transition
(SPT). And significantly a RBM-like mode appears at about 509
cm$^{-1}$, which could be considered as a fingerprint of the SPT
happened in the SWNT bundle, and further used to determine the
microscopic structure of the bundle after the SPT.

\end{abstract}


\maketitle

The carbon nanotubes (CNTs) have attracted much attention since
their discovery [1] due to their remarkable electronic and
mechanical properties [2], and potential applications in future
nanoscale electronic devices. It is well known that the physical
properties of the CNTs depend much on their geometrical
structures, and so can be easily changed by an applied pressure or
strain, which could be used to fabricate the nanoscale
electromechanical coupling devices and transducers. For example, a
uniaxial strain on the SWNTs can cause a metal-semiconductor
transition [3].

Recently, the effect of hydrostatic pressure on the CNT bundle,
including the Raman spectroscopy investigation, has attracted much
attention in experiments [4-10] and theoretical calculations
[11-15]. It is reported that the intensity and broadening of $R$
band decrease with increasing pressure and the RBM vanishes above
a critical pressure [4-7], showing a SPT at this pressure.
Meantime, the Raman spectra [4], $in$ $situ$ synchrotron X-ray
diffraction data [8] and the optical absorption spectra [9] showed
that the changes of geometrical structure and optical properties
are reversible upon unloading the pressure. Sluiter and Kawazoe
[11] used the continuum model to calculate the complete phase
diagram of the carbon nanotube bundle under hydrostatic pressure.
For the (10,10) SWNT bundle, their results indicate, if
considering stable states only, tube's cross sections are round up
to 1.9 GPa, racetrack shaped between 1.9 and 6.0 GPa. But until
now, how to verify their predictions experimentally is still an
open question, meanwhile no detailed theoretical analysis on the
vibrational properties and Raman spectra of CNT bundle under
hydrostatic pressure has been made, which is just what we attempt
to do in this paper.

The Raman spectroscopy has been considered as a powerful tool to
detect the diameter-selective phonon modes of the SWNTs [16-18].
An important low-frequency Raman peak is attributed to the RBM
($R$ band), where all carbon atoms are subject to an in-phase
radial displacements. It was found that the RBM frequency is
inversely proportional to the tube diameter, and independent of
its chirality. And the higher frequency peaks are caused by the
out-of-phase vibrations of the neighboring carbon atoms parallel
to the surface of the cylinder (tangential modes, $T$ band).

In this work we will pay our main attention to the effect of
hydrostatic pressure on the vibrational modes of the (10,10) SWNT
bundle. Our most striking finding is a RBM-like mode in the
collapsed bundle after the SPT, which is also Raman active and
could be considered as a fingerprint of the SPT happened in the
SWNT bundle, and further used to accurately measure structure of
the bundle after the SPT.

Our theoretical calculations were performed using the
density-functional theory in the local-density approximation (LDA)
[19], in which the highly accurate projected augmented wave (PAW)
method [20] was used. The maximum spacing between K points is 0.03
\AA $^{-1}$ and the Gaussian smearing width is 0.04 eV. The
geometrical structures of the SWNT bundle before and after the SPT
are optimized by the first-principle method at 0, 0.45, 2.8 and 5
GPa, and in the final geometry no forces on the atoms exceed 0.001
eV/\AA. The Raman intensity was calculated by combining above
first-principles results with the empirical bond polarizability
model [21].

\begin{figure}[htbp]
\includegraphics[width=0.8\columnwidth]{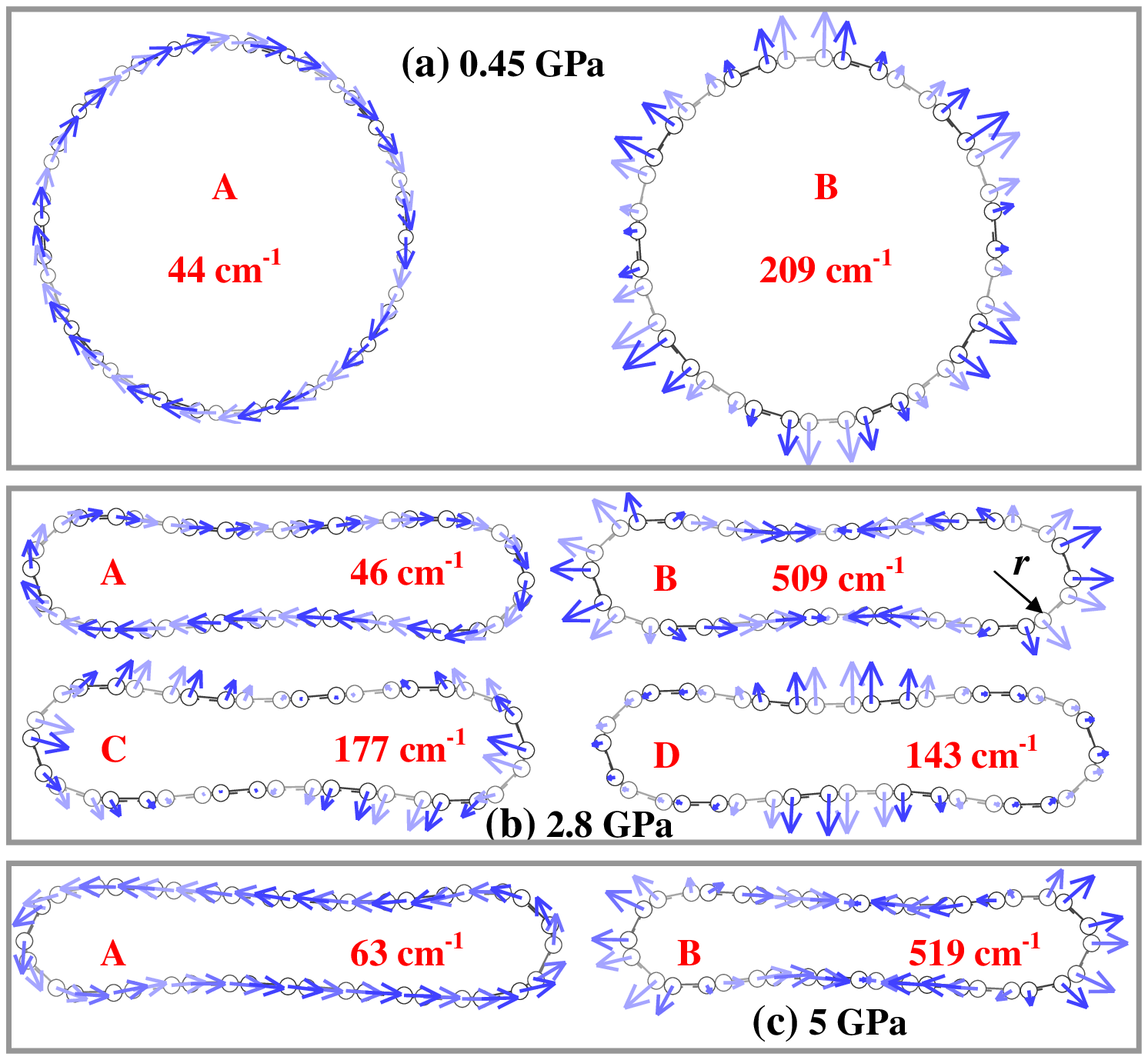}
\label{fig1} \caption{(Color online) Some characteristic vibrational modes of
the bundle, (a) at 0.45 GPa before the SPT, (b) at 2.8 GPa after
the SPT, and (c) at 5 GPa. The small circles represent the carbon
atoms and the straight lines between them indicate the bonds. The
arrows represent the atomic motions. Deeper color arrow means the
corresponding atom is nearer to the reader. $r$ is the radius of
its two circular ends.}
\end{figure}

The optimized structures of a (10, 10) SWNT bundle under different
pressures are shown in Figs. 1(a)-1(c). As the hydrostatic
pressure increases, the tube surface deformation energy increases,
and the intertube distance decreases, increasing the van der Waals
(vdW) energy. When the pressure increases up to a critical value,
it becomes energetically favorable to reduce the tube's inner
volume, rather than simply reduce the intertube distance, making
thus the system undergo a SPT and the tube spontaneously collapses
to form a peanut-like cross-section, which is consistent with the
recent molecular dynamics simulations [14] and continuum model
analysis [11]. Such a kind of structural change should exhibit
some important characters in the vibrational properties.

To investigate the systematic vibrational properties, the
nonresonant Raman spectra of SWNT bundle under different pressures
are firstly calculated and plotted in Fig. 2. It can be found from
Fig. 2 that more peaks emerge in both the low- and high-frequency
regions after the SPT, i.e., the whole Raman spectra distribute
wider after the SPT. This should be attributed to the
symmetry-breaking and the split of phonon bands. The high
frequency part in the Raman spectra of the bundle before the SPT
are almost the same, and the same holds for those after the SPT.
This means the sudden change of structure occurs only at the SPT
pressure. Most importantly, the original RBM of the bundle
disappears from the spectra after the SPT, which is consistent
with the experimental results [5,6].

\begin{figure}[htbp]
\includegraphics[height=1.2\columnwidth]{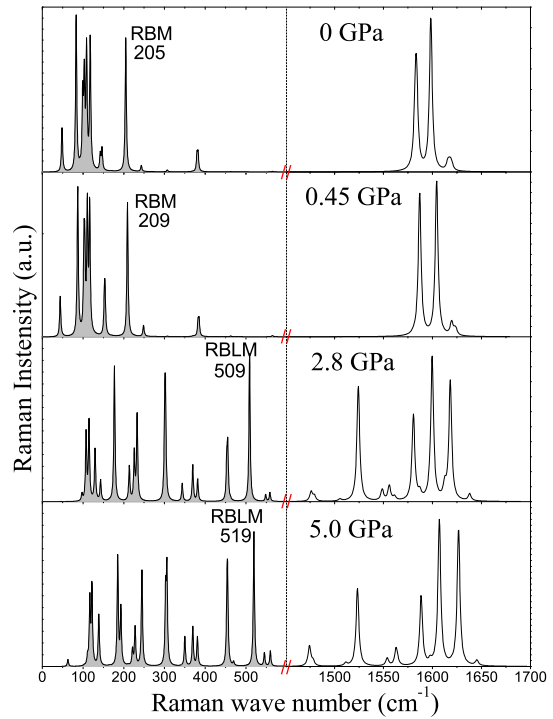}
\label{fig1} \caption{(Color online) The nonresonant Raman spectra of (10, 10)
bundle at different pressures. From up to down, the external
pressure is: 0 , 0.45, 2.8 and 5 GPa. The frequencies of RBM and
RBM-like modes are labeled.}
\end{figure}

In all the structures, the Raman frequencies increase when
external pressure increases. We calculate the pressure dependence
and summarize them in Table I, in which the experimental data
taken from Ref. 5 are also presented. From Table I, we can find
that although first-principle calculation overestimates the Raman
frequency systematically, it can reproduce most of the
experimental dependence of the Raman frequency on hydrostatic
pressure. So we can use the first-principle results to further
investigate in detail the vibration modes of (10, 10) bundle under
hydrostatic pressure.

\begin{table}[htbp]
\caption{\label{table1} The frequencies of some characteristic
Raman active modes and their pressure dependence. The $\omega
_{exp} $ and $\omega _{the} $ are the experimental and theoretical
Raman frequencies of the structure at 0 GPa.}
\begin{ruledtabular}
\begin{tabular}{ccccccccc}
Structure & \multicolumn{4}{c} {RBM} &
\multicolumn{4}{c}{$T$ Band}  \\
region & $\omega _{exp}$&$\frac{d\omega _{exp} }{dp}$& $\omega
_{the} $& $\frac{d\omega _{the} }{dp}$& $\omega _{exp} $&
$\frac{d\omega _{exp} }{dp}$& $\omega _{the} $&
$\frac{d\omega _{the} }{dp}$ \\
\hline Before &
 &
 &
 &
 &
1566& 6.1& 1583&9.3\\
SPT&181& 10.1& 205& 9.0& 1571& 11.0& 1598&
13.0 \\
 &
 &
 &
 &
 &
1591& 10.1& 1615&
9.3 \\
\hline After& & & & & & 0.7& &
1.6 \\
SPT
 &
 &
 &
 &
 &
 &
5.9&
 &
6.2 \\
 &
 &
 &
 &
 &
 &
5.8&
 &
3.9 \\
\end{tabular}
\end{ruledtabular}
\label{tab1}
\end{table}

Because of the complexity of the vibrational modes, here we only
pay main attention to some characteristic vibrational modes in low
frequency region. We show the rotational mode (A), and the RBM or
RBM-like mode (B) of the SWNT bundle at 0.45, 2.8 and 5 GPa in
Figs. 1(a)-1(c) respectively. Firstly, in Fig. 1(a) one may notice
that the tube's cross-section is not a perfect circle now. This is
because the rotational symmetry of the (10,10) tube is
incompatible with the hexagonal lattice symmetry, making the
tube's cross-section deformed slightly. Secondly, the atomic
motions in the rotational mode (A) before and after the SPT are
always almost along the tube surface, but its frequency can not
remain zero because of the non-circular cross-section and the
tube-tube interaction. This mode is an IR active, making it
observed possibly in IR spectra.

On the other hand, because of the hexagonal symmetry of the unit
cell in the bundle, the atomic motions in the RBM are not the same
for every atom, as seen from Fig. 1(a), exhibiting approximately
the hexagonal symmetry too, which means that the RBM frequency in
the SWNT bundle will not be the same as that in the isolated SWNT.

A special attention should be paid to the mode (B) of 509 cm$^{ -
1}$ at 2.8 GPa after the SPT in Fig. 1(b), in which most of the
atomic motions on the two ends are perpendicular to the tube
surface, showing in-phase vibrations, and the motions of the atoms
in the flat region are along the tube surface. This mode also
exists at 5 GPa, as shown in Fig. 1(c), which so can be regarded
as a RBM-like mode appearing only in the bundle after the SPT. In
fact, the previous researcher also found a similar mode in a
capped SWNT [22], which comes from a mixing of the RBM of the
$C_{60}$ hemisphere and the tangential mode of capped SWNT.
Because most of the atomic motions in the mode are in-phase, this
RBM-like mode have a large Raman intensity as shown in Fig. 2,
making it can be found experimentally in the Raman spectra of the
bundle after the SPT.

It is well known that the RBM frequency is inversely proportional
to tube diameter, expressed by the formula of $f_{RBM} = D / d_t $
, where $D$ is a constant. To examine further the RBM-like
behavior of the mode in Fig. 1(b), we have measured the radius of
the circular end in Fig. 1(b), which is $r = \frac{d_t }{2}
\approx 2.26$ {\AA}. After taking $D = 228$ nm/cm$^{ - 1}$ from
Ref. [5], which was also obtained by the first-principle method,
one can obtain $f_{RBM}=\mbox{504.4}$ cm$^{ - 1}$. This value is
very close to our first-principle result (509 cm$^{ - 1}$), and a
difference of several cm$^{ - 1}$ comes from the tube-tube
interactions. In order to make a more accurate comparison between
them, we directly take one isolated tube from the collapsed bundle
at 2.8 GPa, and then calculate the frequency of its RBM-like mode,
which is found to be 501 cm$^{ - 1}$. Thus, a very good agreement
between the normal mode frequency of the isolated tube and the RBM
of an equivalent SWNT composed of the circular ends of a collapsed
tube in the bundle has been obtained, indicating the normal mode
(B) in Fig. 1(b) can indeed be considered as a RBM-like mode of
the tube bundle after the SPT. It is valuable to mention that
appearance of this RBM-like mode indicates an existence of the two
circular ends in the bundle after the SPT, which furthermore can
be regarded as a fingerprint-type mark of the SPT. More
importantly, the frequency of this RBM-like mode is only
determined by the radii of the two circular ends, and so is mainly
influenced by the external pressure, but not the radius of the
original tube. As shown in Fig. 1(c), when external pressure
increases to 5 GPa, the frequency of the RBM-like mode is changed
to 519 cm$^{-1}$. So this mode can also be used to determine
experimentally the microscopic structure of the bundle after the
SPT, no matter what kind of SWNT bundle is used originally.

Another interesting Raman-active normal mode (C) in Fig. 1(b) is a
quadrupole vibration mode, which often has a strong intensity in
the Raman spectra of clusters, and the lowest $E_{2g}$ Raman
active mode of the SWNT is also a quadrupole mode. The similitude
between these modes may indicate that the mode (C) origins from
the lowest $E_{2g}$ Raman active mode of the SWNT. In other words,
the $E_{2g}$-like modes of the two circular ends would combine
together to give the mode (C) in Fig. 1(b), which could be
identified by the following argument. We know that the lowest
$E_{2g}$ Raman active mode of SWNT is inversely proportional to
the squared tube diameter, i.e., $f \propto d_t^{ - 2} $ [23].
Taking $r = \frac{d_t }{2} \approx 2.26$ {\AA}, its $E_{2g}$ mode
frequency will become $f \approx \left( {\frac{6.8}{2.26}}
\right)^2\times 17 \approx \mbox{154}$ cm$^{-1}$, where the radius
of (10, 10) tube is equal to $r_{(10,10)} \approx 6.8$ \AA, and 17
cm$^{-1}$ is the $E_{2g}$ mode frequency of the isolated (10, 10)
SWNT. After considering the up-shift of the frequency induced by
the intertube interaction, this value is compatible with the mode
frequency of 177 cm$^{-1}$ given in Fig. 1(b). So, we think this
mode can also be observed in future Raman experiments.

Next, the mode (D) of 143 cm$^{-1}$ in Fig. 1(b) is mainly
contributed by the out-phase motion of nearby flat walls, which
can be regarded as the $B_{2g}$ mode of the graphite with its
frequency of 127 cm$^{-1}$. But now due to pressure effect, its
frequency increases to 143 cm$^{-1}$. When pressure increases up
to 5 GPa, the frequency of this mode further increases to 193
cm$^{-1}$.

In conclusion, using the first principles calculations we have
studied in detail the vibrational modes of the (10, 10) SWNT
bundle under hydrostatic pressure, which are found to be very
different from those without the applied pressure, especially
after the SPT. Several important results are obtained, e.g.,
disappearance of the original RBM, and appearance of a RBM-like
mode and a quadrupole vibrational mode after the SPT. All the
modes can be considered as a fingerprint of happening of the SPT,
and used to measure accurately the microscopic structure of the
bundle after the SPT in future experiment. On the other hand, the
characters mentioned above have a close relationship with the
peanut-like cross-section, and so our results show that Raman
spectra can be used to verify the prediction of the foregone
researches [11, 14].

Acknowledgments: The authors acknowledge support from the Natural
Science Foundation of China under Grant No. 10474035 and No.
A040108, and also support from a Grant for State Key Program of
China through Grant No. 2004CB619004.

\end{document}